\documentclass[notitlepage,superscriptaddress,showpacs,nobalancelastpage,twocolumn,aps,prl]{revtex4-1}
\usepackage[english]{babel}
\RequirePackage[T1]{fontenc}
\RequirePackage{times} 

\usepackage{amsfonts}
\usepackage{subfigure}
\usepackage{amsmath}
\usepackage{amssymb}
\usepackage{bm}
\usepackage{graphicx,epstopdf}
\usepackage{appendix}
\usepackage{array}
\usepackage{color}
\usepackage{cancel}
\usepackage{my_symbols}
\usepackage[normalem]{ulem}
\usepackage{CJK}
\providecommand{\U}[1]{\protect\rule{.1in}{.1in}}

{\par}

\newcommand\rmv{\bgroup\markoverwith {\textcolor{red}{\rule[0.5ex]{2pt}{0.4pt}}}\ULon}

\begin{document}

\begin{CJK*}{UTF8}{gbsn} 
\title{Scaling of anomalous Hall effect in ferromagnetic thin films}
\author{Vahram L. Grigoryan}
\affiliation{Department of Physics, Beijing Normal University, Beijing 100875, China}
\author{Jiang Xiao (萧江)}
\email[Corresponding author:~]{xiaojiang@fudan.edu.cn}
\affiliation{Department of Physics and State Key Laboratory of Surface Physics, Fudan University, Shanghai 200433, China}
\affiliation{Collaborative Innovation Center of Advanced Microstructures, Nanjing, 210093, China}
\author{Xuhui Wang}
\email[Current address:~]{Kwantum Links, Benoordenhoutseweg 23, 2596 BA, The Hague, The Netherlands} 
\affiliation{King Abdullah University of Science and Technology (KAUST), Physical Science and Engineering Division, Thuwal 23955-6900, Saudi Arabia}
\author{Ke Xia}
\email[Corresponding author:~]{kexia@bnu.edu.cn}
\affiliation{Department of Physics, Beijing Normal University, Beijing 100875, China}

\begin{abstract}

We propose a new scaling law for anomalous Hall effect in ferromagnetic thin films by distinguishing three scattering sources, namely, bulk impurity, phonon, and more importantly a rough surface. This new scaling law fits the recent experimental data excellently with constant coefficients that are independent of temperature and film thickness. This is in stark constrast with previous scaling laws that use temperature/thickness dependent fitting coefficients, and is a strong indicator that this law captures the essential physics. By intepretating the experiments for Fe, Co, and Ni with this new law, we conclude that (i) the phonon-induced skew scattering is unimportant as expected; (ii) contribution from the impurity-induced skew scattering is negative; (iii) the intrinsic (extrinsic) mechanism dominates in Fe (Co), and both the extrinsic and the intrinsic contribution are important in Ni.


\end{abstract}
\pacs{}
\maketitle
\end{CJK*}

{\it Introduction.} In normal metals, the Hall effect, a charge current transverse to both the applied magnetic and electric field, is known to be driven by the Lorentz force. In ferromagnetic metals, yet a very similar effect - known as the anomalous Hall effect (AHE) - is present without magnetic field \cite{pugh_hall_1953, nagaosa_anomalous_2010}.  The real physical mechanisms behind AHE are under debate for decades. Spin-orbit coupling (SOC) - believed to be responsible for AHE - brings two types of mechanisms, i.e., the intrinsic and the extrinsic one \cite{nagaosa_anomalous_2010}. The intrinsic mechanism, arising from the SOC intrinsic to the band structure, was first proposed by Karplus and Luttinger \cite{karplus_hall_1954} and later reformulated in terms of Berry's phase \cite{sundaram_wave-packet_1999}. The extrinsic mechanism, due to scatterings with impurities carrying SOC, gives rise to two contributions known as skew-scattering \cite{smit_spontaneous_1955,smit_spontaneous_1958} and side-jump \cite{berger_side-jump_1970}. Asymmetric scattering by impurities leads to the skew scattering, while the side-jump originates from that electrons with opposite spins are deflected to opposite directions when scattered by an impurity.

The progress made in the past indicates that the intrinsic mechanism shall dominate in AHE \cite{miyasato_2007_int-dom-ext}.  The theories suggest that the Hall resistivity ($\rho_\ssf{AH}$) and longitudinal resistivity $\rho$ has the following scaling relations: $\rho_\ssf{AH}^\ssf{INT}, \rho_\ssf{AH}^\ssf{SJ} \propto \rho^2$ for both the intrinsic \cite{karplus_hall_1954} and extrinsic side-jump mechanism \cite{berger_side-jump_1970}, while $\rho_\ssf{AH}^\ssf{SS}\propto \rho$ for the extrinsic skew-scattering mechanism \cite{smit_spontaneous_1955,smit_spontaneous_1958}. Such a scaling law is helpful to outline the underlying mechanism. But, this simple relation $\rho_\ssf{AH} \sim a \rho + b \rho^2$ often breaks down when a comparison with experiments is made.

To resolve this issue \cite{onoda_2006-int-vs-ext}, Jin's group in Fudan University - by systematically varying the film thickness and temperature -  has measured both the longitudinal and anomalous Hall resistivities in Fe, Co, and Ni thin films \cite{tian_proper_2009,hou_anomalous_2012,ye_temperature_2012,hou_multivariable_2015}.  Such an experimental paradigm does not affect sample's band structure and leaves the intrinsic contribution untouched. Hou \etal \cite{hou_multivariable_2015} argued that, while most theories only assumes a single type of scatter, the real complexity in the experimental data arises from the fact that there are more than one type of impurity scatters. By assuming multiple scatters, Hou \etal proposed a scaling law \cite{hou_multivariable_2015}
\begin{equation}
\rho_\ssf{AH}=
c\rho_{xx}^2 + \sum_i c_i\rho_i\rho_\ssf{xx} + \sum_{ij}c_{ij}\rho_i\rho_j + \sum_{i\in S}\alpha_i\rho_i
\label{eqn:houscaling}
\end{equation}
where $\rho_{xx} = \sum_i\rho_i$ is the overall longitudinal resistivity, $\rho_i$ the partial longitudinal resistivity from $i$-th scattering source, and $S$ represents the static scattering sources that remains at zero temperature. And $c, c_i, c_{ij}, \alpha_i$ are constant coefficients. We believe that this new relation points to a plausible direction for resolving the AHE scaling issue. Unfortunately, this scaling law not only carries extreme complexity but also is very difficult to test since the partial resistivities are usually not directly measurable.


In this Letter, we develop our theory to describe the AHE in ferromagnetic thin films. We calculate the longitudinal resistivity $\rho$, transverse resistivity $\rho_\ssf{AH}$, and thus the scaling relation by taking into account the finite thickness and scattering by the bulk impurities, phonons and surface roughness. We find excellent agreement between our theory and the experimental data for Fe, Co, and Ni obtained by Jin's group \cite{tian_proper_2009,hou_anomalous_2012,ye_temperature_2012,hou_multivariable_2015}. For $\rho$, the film roughness is the only fitting parameter; for $\rho_\ssf{AH}$, there are only four constant coefficients accounting the impurity-induced side-jump and skew-scattering, phonon-induced side-jump, and intrinsic contributions. We are able to identify the proportion of each contribution. And we conclude that, the intrinsic (extrinsic) effect dominates in Fe (Co). Yet in Ni, the intrinsic and extrinsic are both important and in competition.

{\it Model.} The Hamiltonian for a ferromagnetic thin film with magnetization $\mb$ and a constant thickness $d$:
\begin{equation}
\hH_0=\frac{\hbp^2}{2m}+V_{d}(z) +V_{\rm sd}
\label{eqn:ham}
\end{equation}
where $m$ is the electron mass and $\hbp = -i \hbar \boldsymbol{\nabla}$ momentum operator. The confining potential $V_{d}(z) = U\Theta \smlb{z-d/2}+U\Theta\smlb{-z-d/2}$ with $U$ the potential height and $\Theta$ the Heaviside unit step function. $V_{\rm sd} = -J_{\rm sd} \hbsigma \cdot\mb$ is the exchange energy between the itinerant $s$ and the local $d$ electrons. $J_{\rm sd}$ is s-d coupling constant and the Pauli matrices are $\hbsigma = (\hsigma_x, \hsigma_y, \hsigma_z)$. The eigen-solutions of this unperturbed Hamiltonian $\hH_0$ are
\begin{align}
E_{n \bq s} &= E_{n \bq }-s J_{ \rm s d}, \nn
\ket{\alpha } = \ket{n \bq s} &= \sqrt{{2\ov  Ad}}\sin\smlb{k_nz}e^{i\bq\cdot \brho}\ket{s}
\label{eqn:wf}
\end{align}
where $E_{n\bq}=\hbar^2 (k_n^2 + \bq^2) /2m$ is the kinetic energy with $k_n={\pi n/d}$ denoting different transverse conducting channels. $\brho=(x,y)$ and $\bq=(q_x,q_y)$ are the in-plane coordinate and wavevector, respectively. The area of the film is $A$. Here, $\ket{s} = \ket{\pm}$ is eigenstate of $\hbsigma\cdot\mb\ket{s}=s\ket{s}$.

Now we focus on scattering mechanisms treated as perturbation. First of all, we assume a non-magnetic point-like bulk impurity scatterers with potential $V_\ssf{I}(\br)=V_{\rm imp} \sum_i \delta(\br-\br_i)$ distributed homogeneously in the film. We assume a density $n_i$. Position of impurity-$i$ is given by $\br_i = (\brho_i, z_i)$. Next, we consider electron-phonon interaction with effective potential $V_\ssf{P}.$ Furthermore, the gradient of the impurity (phonon) potential $V_\ssf{I,P}$ gives rise to a spin-orbit (spin-phonon  \cite{spin_phonon_1,spin_phonon_2}) coupling,
\begin{equation}
V^{\rm so}_\ssf{I,P}(\br)={\eta \ov \hbar} \midb{\hbsigma \times \nabla V_\ssf{I,P}(\br)}\cdot \hbp,
\label{eqn:soisd}
\end{equation}
where $\eta = \mu_B \hbar / \smlb{mce}$ is the coupling constant and $\mu_B=e \hbar/\smlb{2 m c}$ is Bohr magneton. In addition, we consider a rough surface described by a fluctuating position-dependent thickness $d(\brho)$ with an average $\avg{d(\brho)} = d$. The surface roughness is converted into an effective scattering potential \cite{tesanovic_quantum_1986,trivedi_quantum_1988,zhou_spin_2015}
\begin{equation}
V_\ssf{R}(\br)= \lambda_{\brho}\midb{2V_{d}(\br)+z \partial_z V_{d}(\br)},
\label{eqn:rough}
\end{equation}
by a dilation operator that transforms a varying thickness into a constant one \cite{tesanovic_quantum_1986}. Here, $\lambda_{\brho}=\ln[d/d(\brho)]$ is the small deviation of the thickness from $d$. We further treat it simply as a "white noise" surface profile, \ie the surface roughness is uncorrelated and characterized by the dimensionless parameter $\Lambda\sim (\delta/d)^2$ with a variance $\delta^2$. This means the correlation $\avg{\lambda_{\brho}\lambda_{\brho'}} = \Lambda a_0^2\delta(\brho-\brho')$, given $a_0=\pi/\kF$. Accounting all contributions above $U = V_\ssf{I} +V_\ssf{P}+ V_\ssf{R} + V_\ssf{I}^{\rm so}+V_\ssf{P}^{\rm so}$, the total Hamiltonian $\hH = \hH_0 + U$.  

We proceed with Lippmann-Schwinger formalism to calculate the transition probability \cite{wang_scaling_2014}
\begin{equation}
P_{n\bq s}^{n'\bq's'}={2\pi \ov \hbar}
\abs{\Avg{\brak{n'\bq's'}{\hat{T}}{n\bq s}}_{\rm en}}^2
\delta\smlb{E_{n\bq s}-E_{n'\bq's'}},
\label{eqn:transition_prob}
\end{equation}
where $\hT = U + U(E-\hH)^{-1}U$.
\begin{figure}[t]
\includegraphics[width=.7\columnwidth]{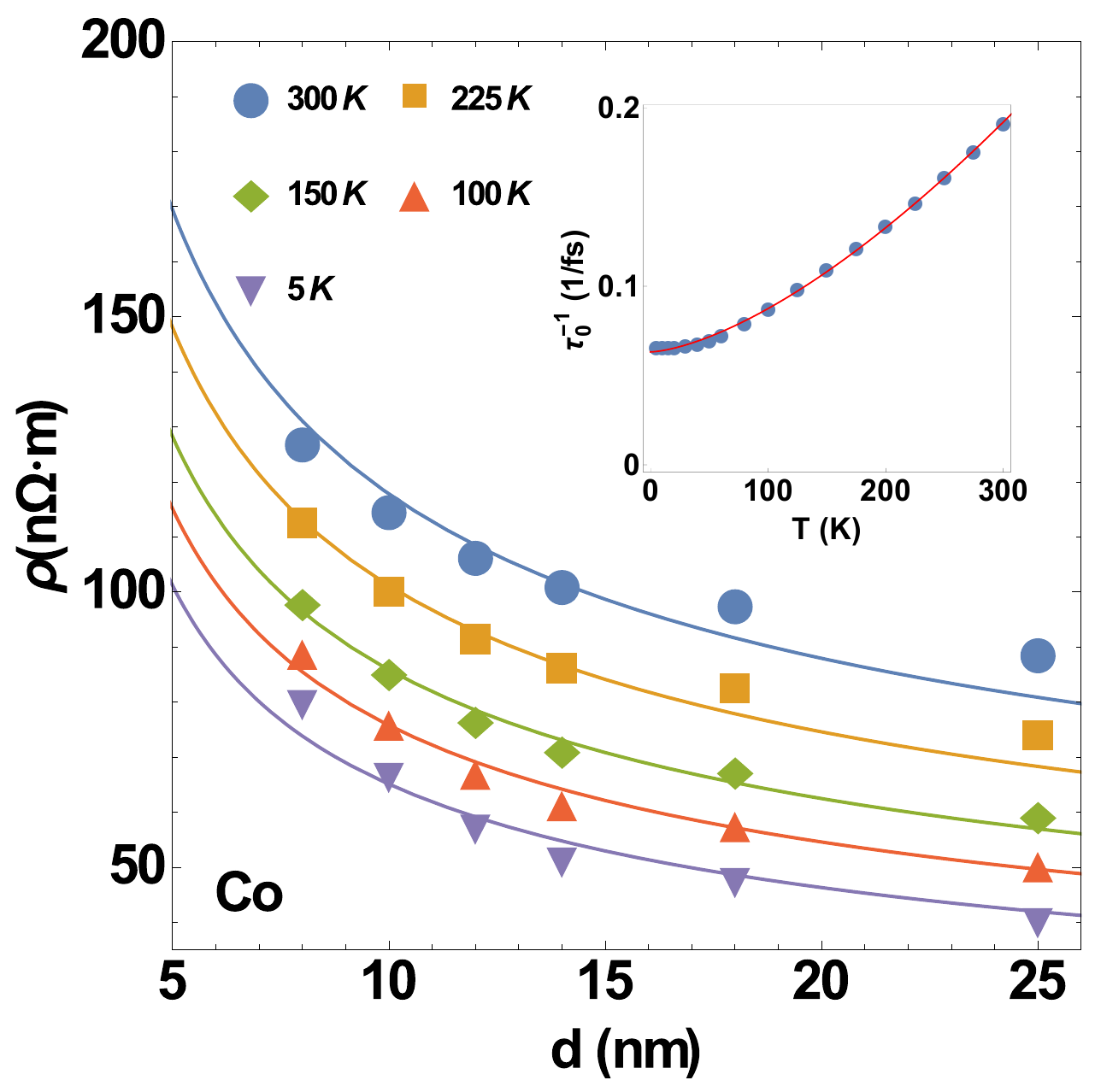}
\caption{(Color online) The thickness dependence of the longitudinal resistivity at various temperatures. Curves at different temperatures have different $\tau_0$, which is plotted in the inset and can be well described by the BG formula (the red curve). The points are the experimental data for Co from Ref. \cite{hou_anomalous_2012}, the curves are plotted from \Eq{eqn:sigxx} with $\delta \approx 5.7~ a_0$.  }
\label{fig:rhoxxslices}
\end{figure}

{\it Longitudinal resistivity.} For each conduction channel at Fermi energy $\EF$, the relaxation rate is obtained from the symmetric part of the transition probability \Eq{eqn:transition_prob} \cite{takahashi_spin_2008,zhou_spin_2015}
\begin{equation}
{1\ov \tau_{n}^s}= {1\ov \tau_{n}}\smlb{1-s{J_{\rm sd}\ov 2 \EF}} \qwith
{1\ov \tau_{n}} = {1\ov \tau_0} + {1\ov \tau^\ssf{R}_n},
\label{eqn:tauns}
\end{equation}
where $\tau_0^{-1}=\tau_\ssf{I}^{-1}+\tau_\ssf{P}^{-1}$ is the bulk relaxation rate subscribing to both the impurity-induced and phonon-induced relaxations. The temperature dependence of $\tau_0^{-1}$ can be approximately described by the Block-Gr\"uneisen (BG) theory, i.e., $\tau_0^{-1} = a+bT^c$ \cite{ziman_electrons_2001,gruneisen_abhangigkeit_1933,bid_temperature_2006}.  And the channel ($n$)-dependent roughness-induced relaxation rate is \cite{tesanovic_quantum_1986,zhou_spin_2015}
\begin{equation}
{1\ov \tau^\ssf{R}_n}=2{\EF\ov\hbar }{ n^2 \ov 3 n_c^3 }\smlb{\delta \ov a_0}^2
\label{eqn:tauR}
\end{equation}
with $n_c = \floor{\kF d/\pi}$ the total number of transverse channels and $n\le n_c$.

\begin{table*}[t]
\caption{Fitted values of the coefficients and surface roughness of thin films.}
\begin{center}
\begin{tabular*}{\textwidth}{@{\extracolsep{\fill}} rlccccccc} \hline\hline
Material & $\kF$~(1/\AA) & $\delta/a_0$ & $\alpha_\ssf{I}~(1/{\rm m}\Omega\cdot{\rm m})$ & $\beta_\ssf{I}~(10^{-3})$ & $\alpha_\ssf{P}~(1/{\rm m}\Omega\cdot{\rm m})$ & $\beta_\ssf{P}$ & $\gamma~(1/{\rm m}\Omega\cdot{\rm m})$ \\ \hline
Fe \cite{hou_multivariable_2015}& $1.71$ \cite{Ashcroft} & $5.2$   & ${+184}$  & ${-7.6}$ & ${-29}$ & $0$ & ${+123}$   \\
Co \cite{hou_anomalous_2012}    & $1.78$ \cite{batallan_1975_kfnico} & $5.7$  & ${+423}$  & ${-9.3}$  & ${+85}$  & $0$ & ${+14}$  \\
Ni \cite{ye_temperature_2012}   & $1.54$ \cite{batallan_1975_kfnico,ehrenreich_1963_kfni} & $4.5$ & ${+728}$  & ${-4.4}$  & ${-47}$  & $0$ & ${+50}$   \\ \hline \hline
\end{tabular*}
\end{center}
\label{tab:param}
\end{table*}
\begin{figure*}[t]
\includegraphics[height=3.7cm]{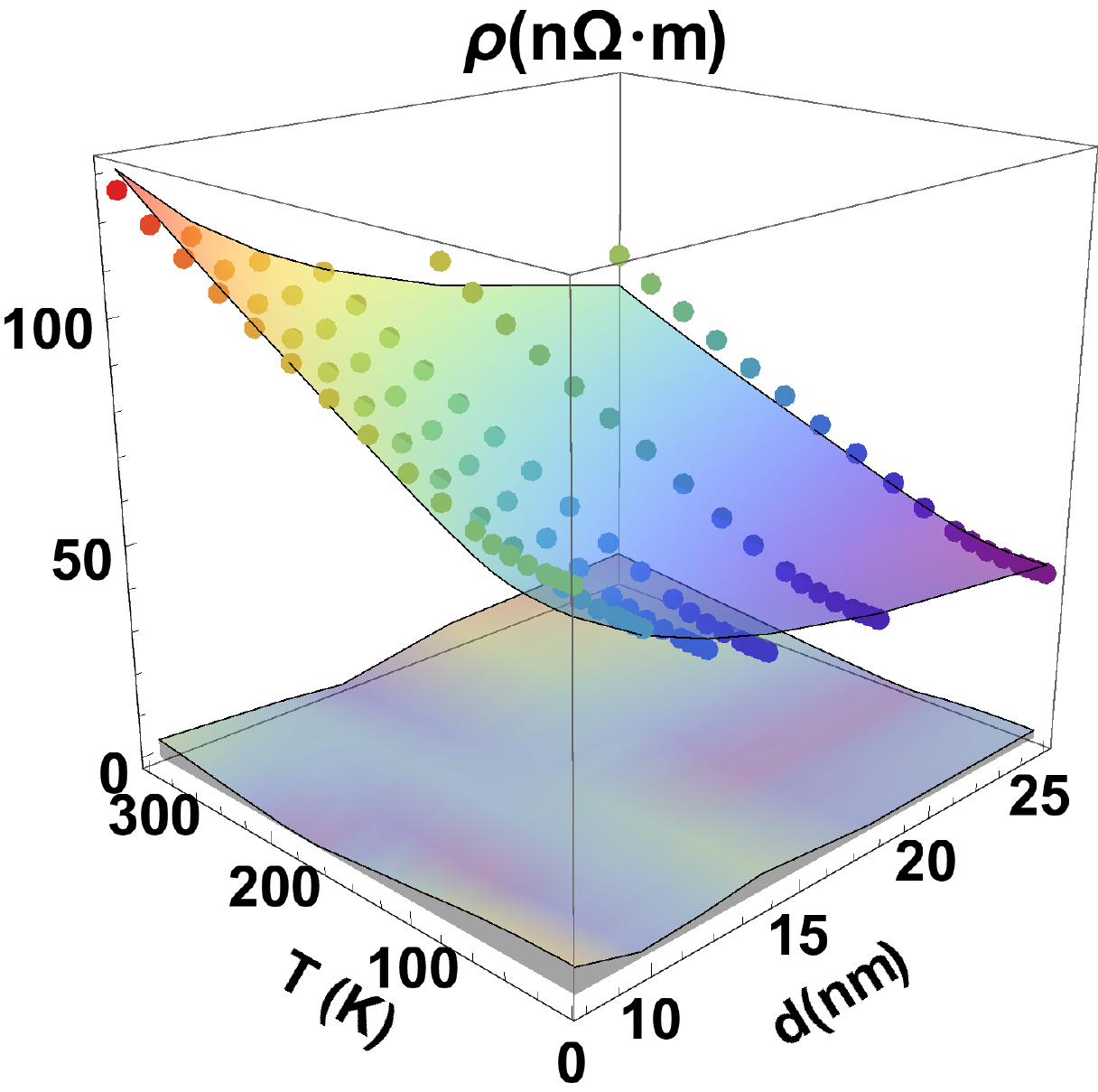}\hspace{0.1cm}
\includegraphics[height=3.7cm]{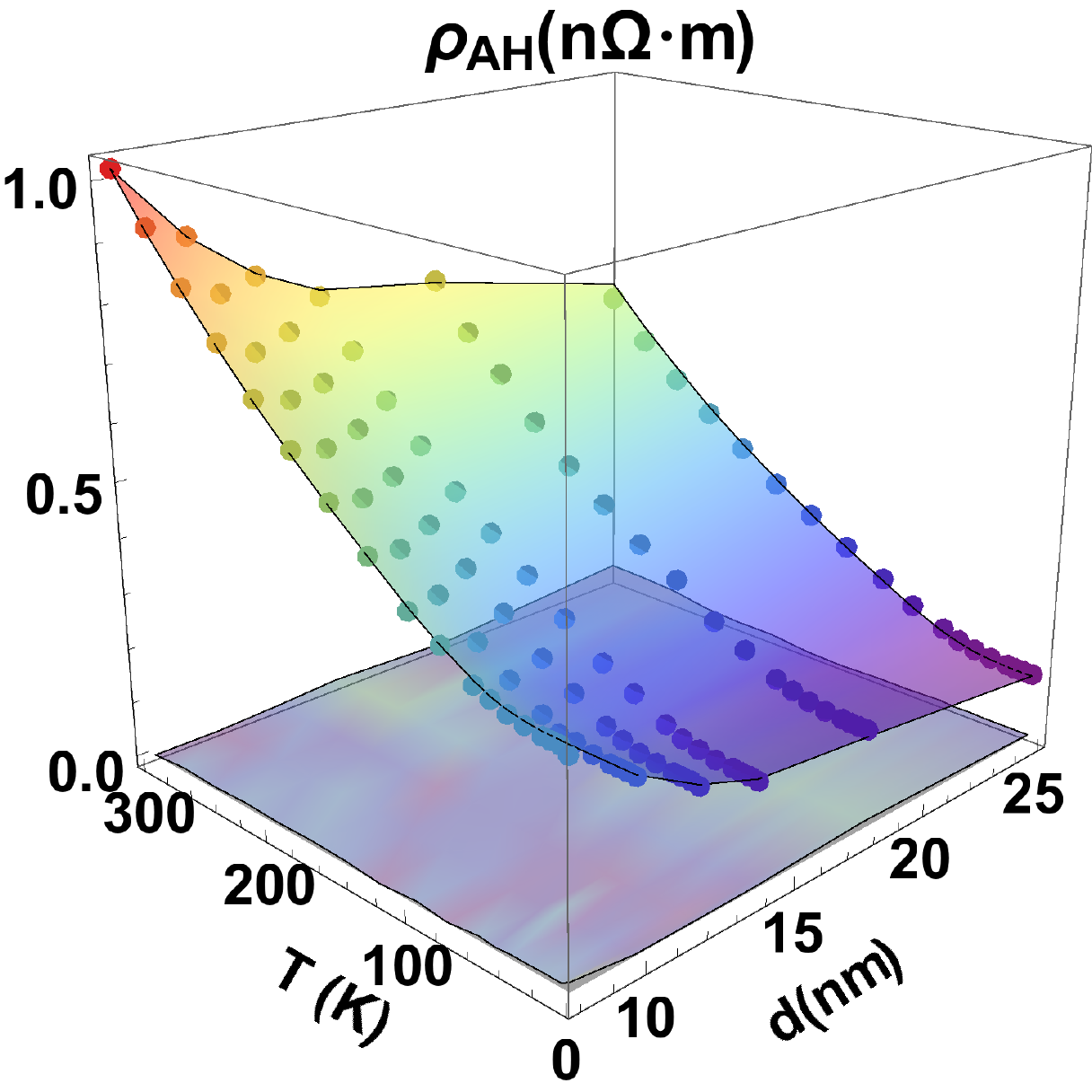}\hspace{0.4cm}
\includegraphics[height=3.7cm,bb=0 10 230 230,clip]{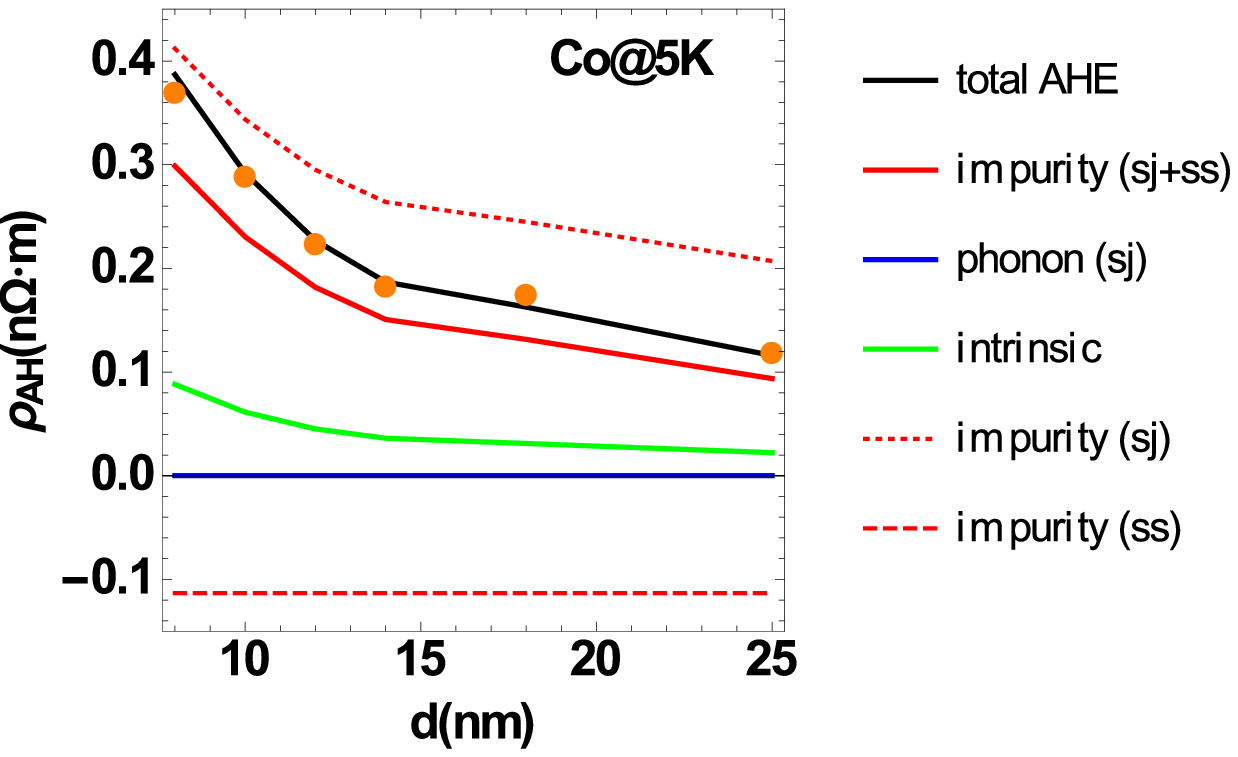} \hspace{0.1cm}
\includegraphics[height=3.7cm,bb=20 10 360 230,clip]{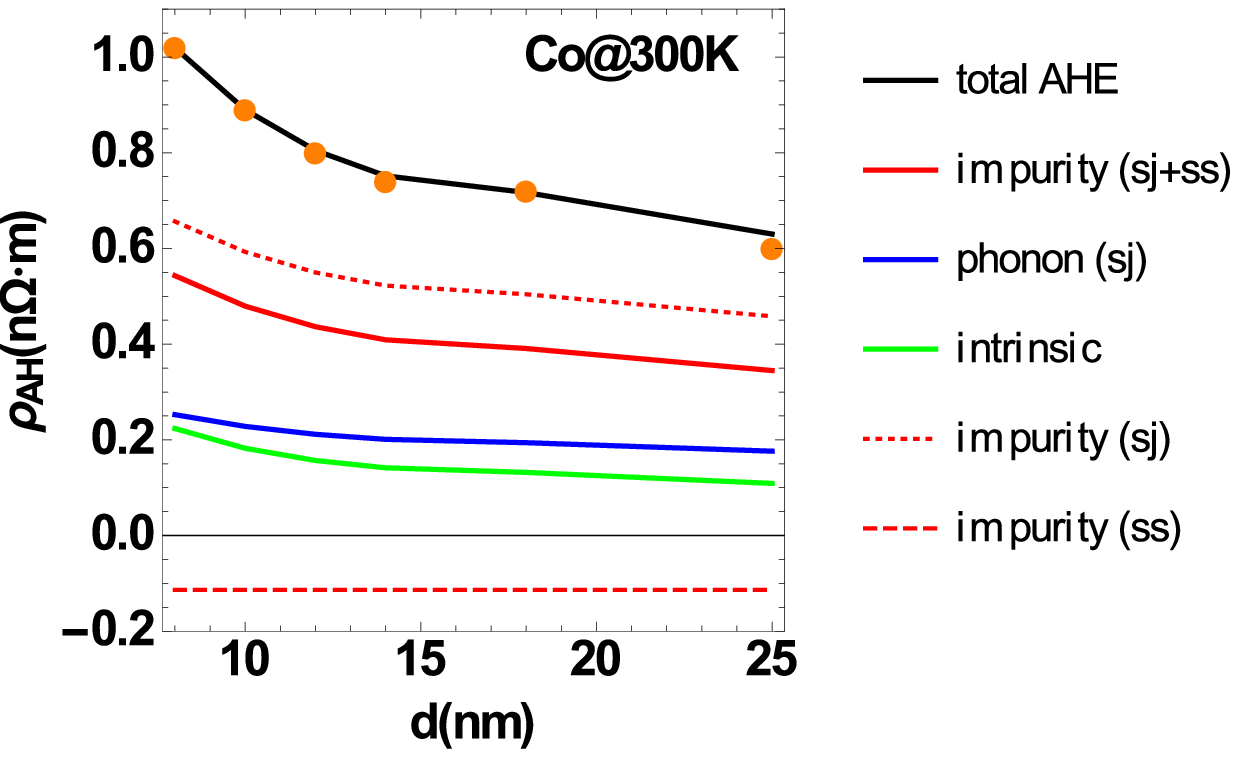}
\caption{(Color online) a) The thickness and temperature dependence of the longitudinal resistivity. The surface is plotted from \Eq{eqn:sigxx}. b) The thickness and temperature dependence of the anomalous Hall resistivity. The surface is plotted from \Eq{eqn:scaling} with $\alpha_{\ssf{I}},$ $\alpha_{\ssf{P}}$, $\gamma$ as fitting parameters, whose values are listed in \Table{tab:param}. The points are the experimental data for Co from Ref. \cite{hou_anomalous_2012}. The shaded area at the bottom indicates the absolute difference between the experiment and theory. c) and d) thickness dependence of extrinsic and intrinsic contributions at low and high temperatures, respectively. The dotted/dashed red lines is the side-jump/skew-scattering contribution due to impurity scattering, and their sum is plotted as the solid red line. The blue solid line is the side-jump contribution due to phonon scattering. The solid green line is the intrinsic contribution. }
\label{fig:FeCo}
\end{figure*}

The relaxation rates in \Eqs{eqn:tauns}{eqn:tauR} let us calculate the in-plane longitudinal conductivity as a function of film thickness, surface roughness, and temperature \cite{tesanovic_quantum_1986,trivedi_quantum_1988,zhou_spin_2015,wang_scaling_2014}
\begin{equation}
\rho^{-1} = \sigma(T,d,\delta) = {3\sigma_0 \ov 2n_c}
\sum_{n}^{n_{c}}{\tau_{n} \ov \tau_0} \smlb{1-{n^2\ov n_{c}^2}},
\label{eqn:sigxx}
\end{equation}
where $\sigma_0= n_ee^2 \tau_0/m$ is the temperature dependent bulk Drude conductivity with an electron density $n_e = \kF^3/3\pi^2$.

\begin{figure*}[t]
\includegraphics[height=3.7cm]{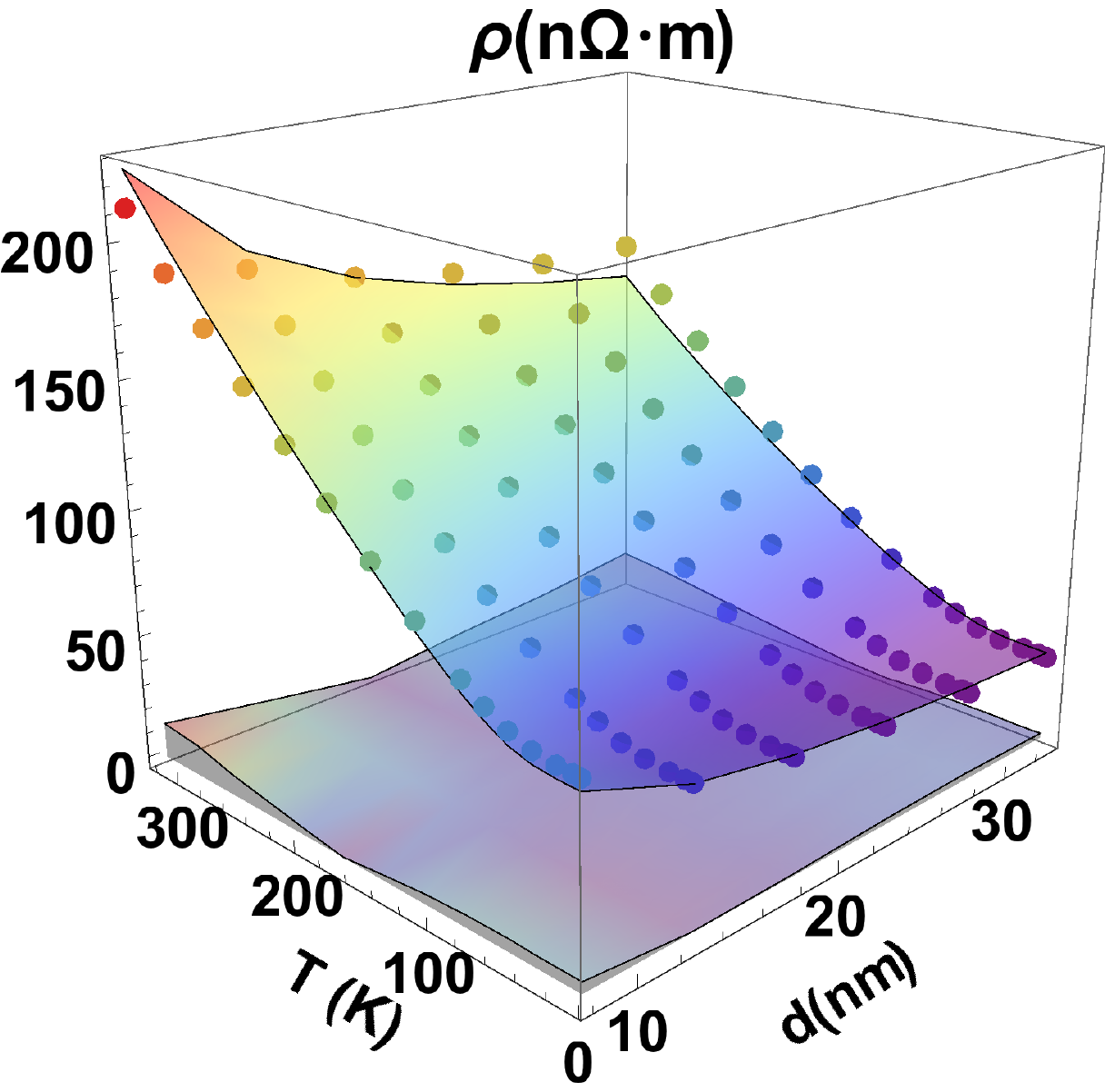}\hspace{0.1cm}
\includegraphics[height=3.7cm]{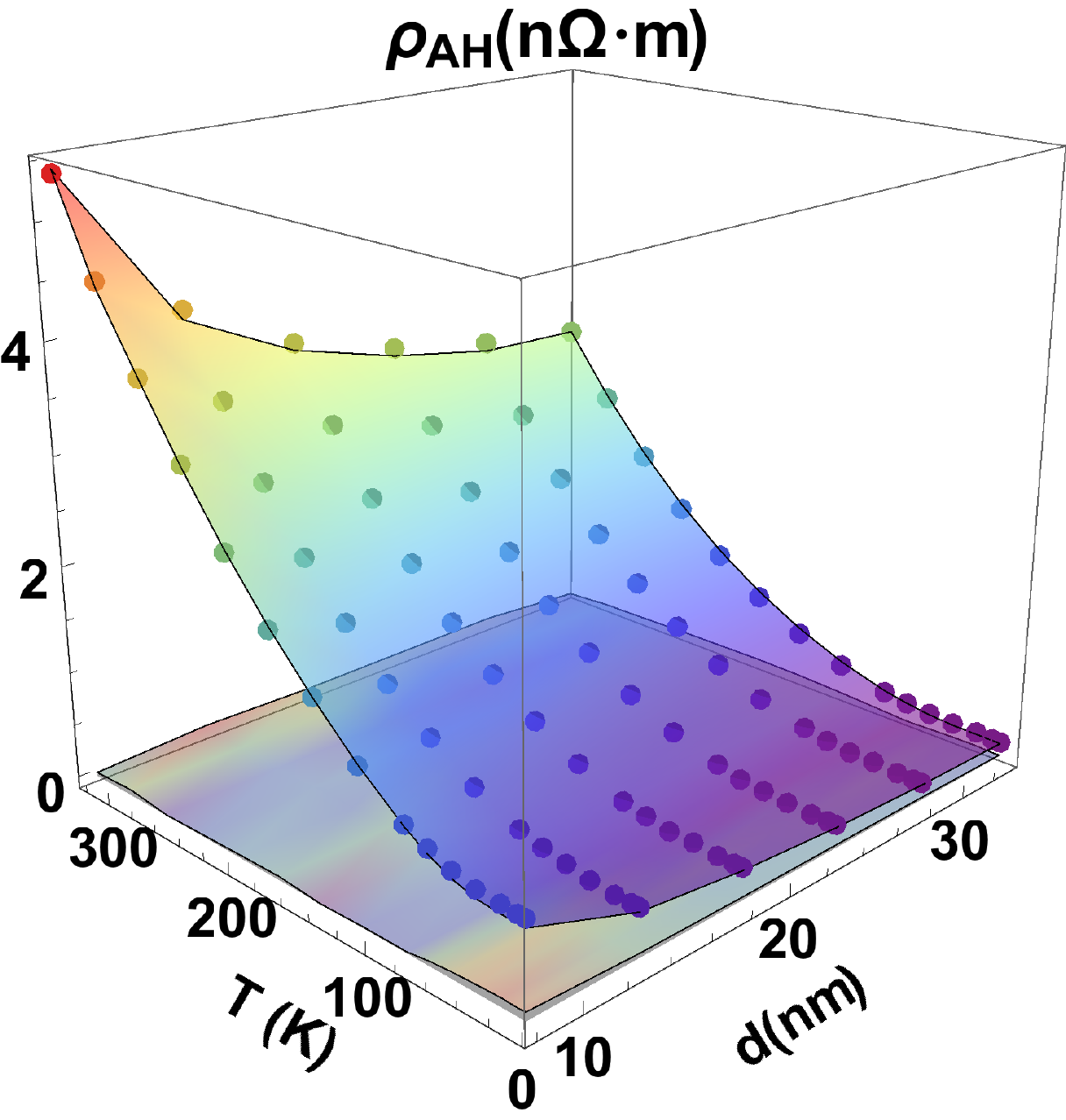}\hspace{0.4cm}
\includegraphics[height=3.7cm,bb=0 10 360 230,clip]{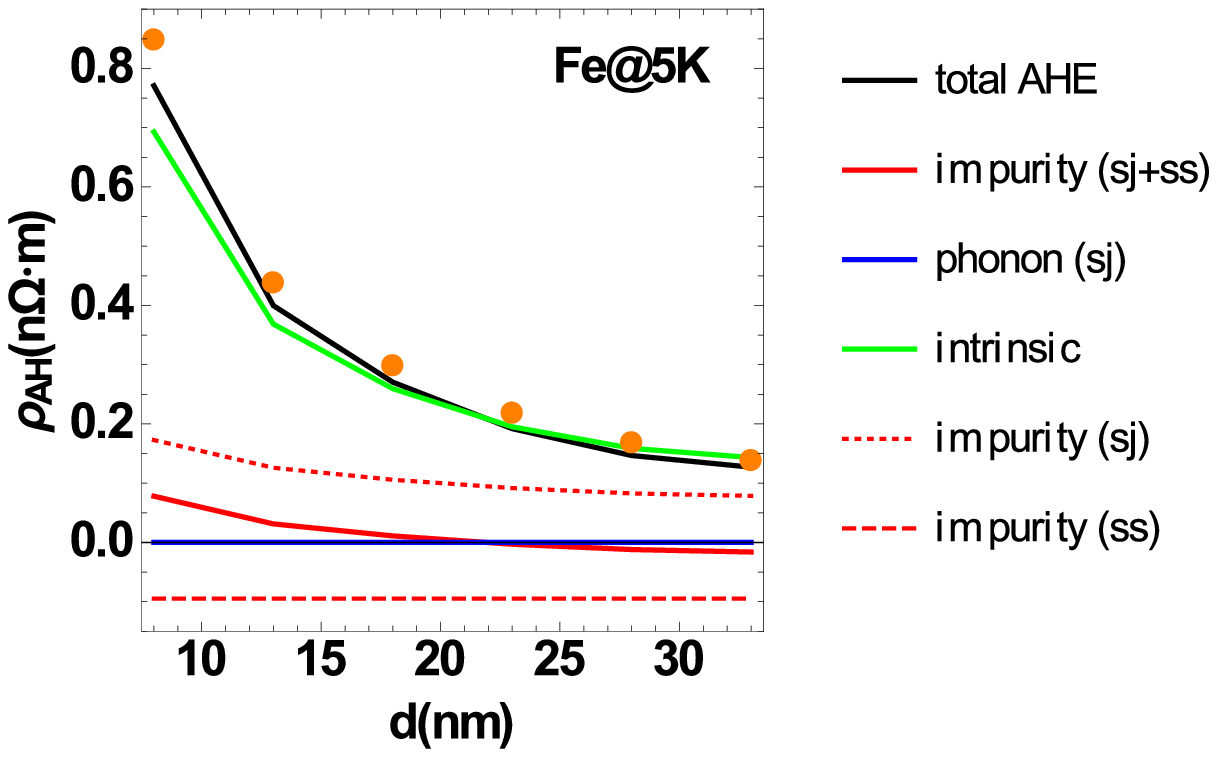}
\includegraphics[height=3.7cm,bb=20 10 230 230,clip]{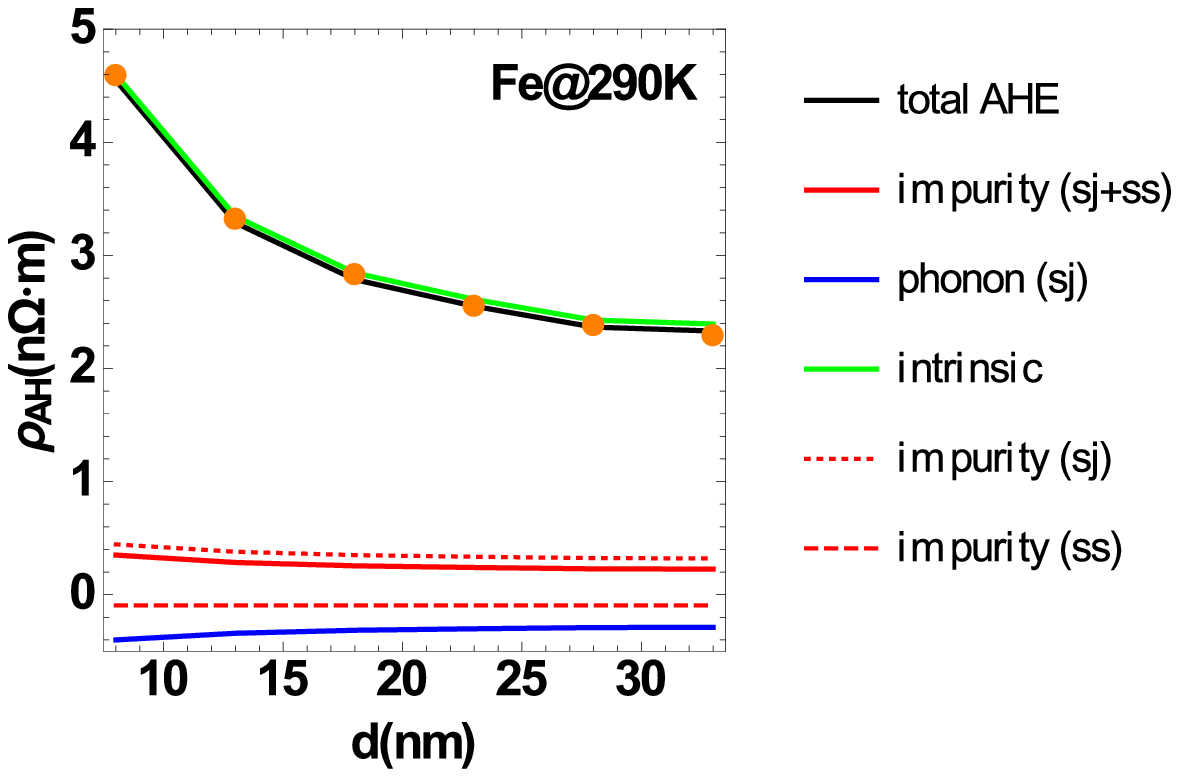}
\caption{(Color online) The same as in \Figure{fig:FeCo} for Fe. The points are the experimental data from Ref. \cite{hou_multivariable_2015}.}
\label{fig:FeFe}
\end{figure*}

\begin{figure*}[t]
\includegraphics[height=3.7cm]{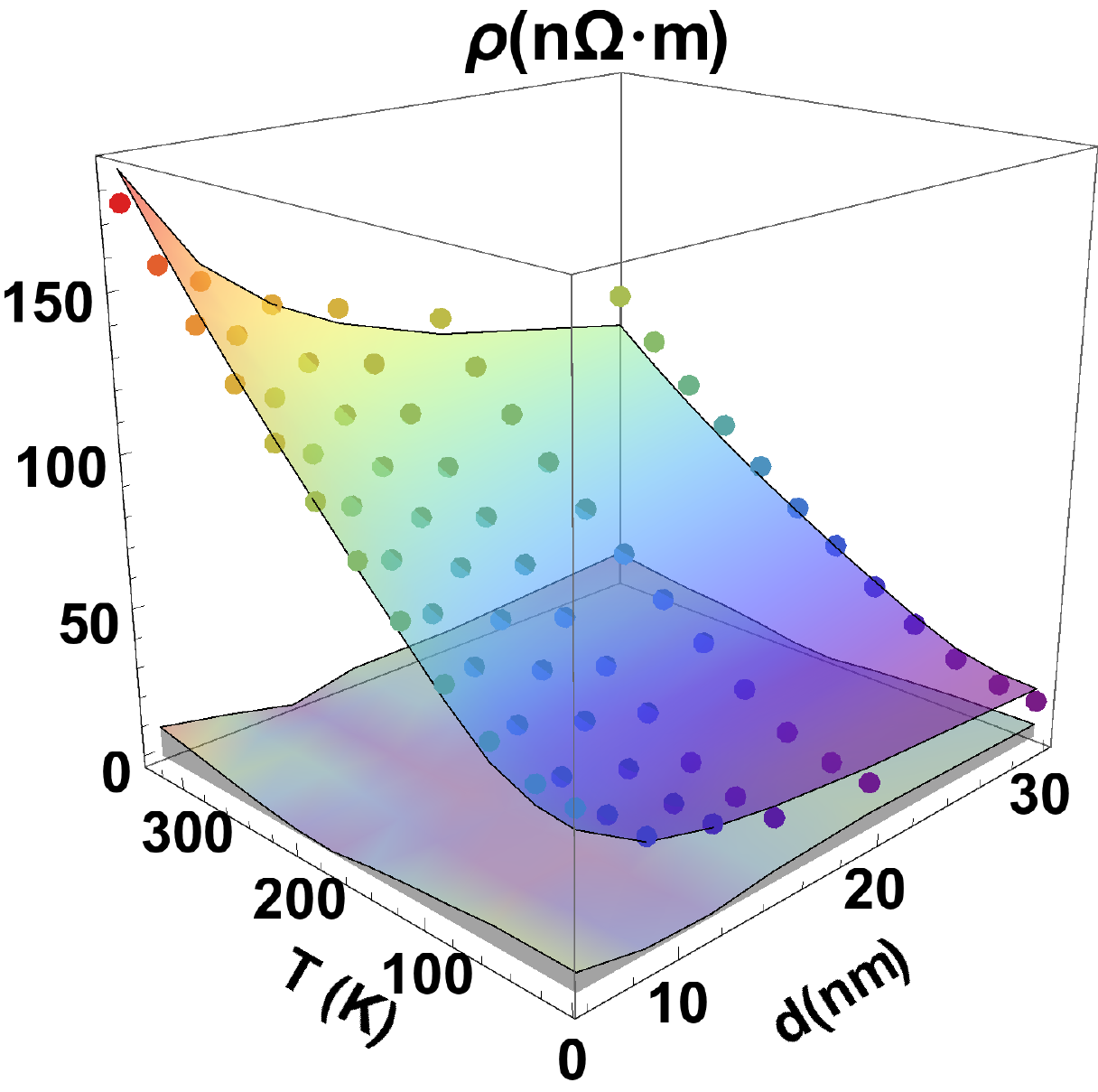}\hspace{0.1cm}
\includegraphics[height=3.7cm]{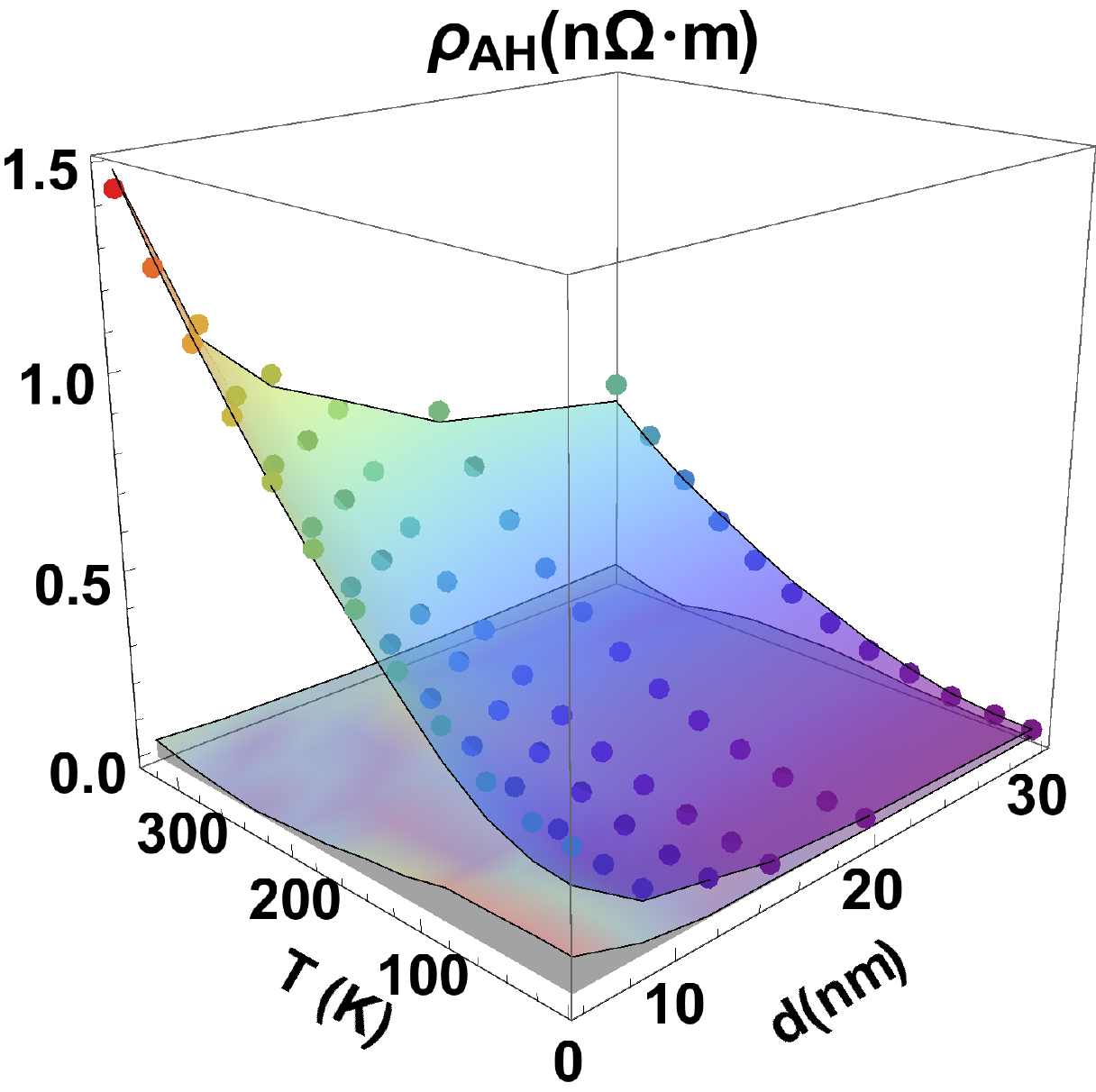}\hspace{0.4cm}
\includegraphics[height=3.7cm,bb=0 10 360 230,clip]{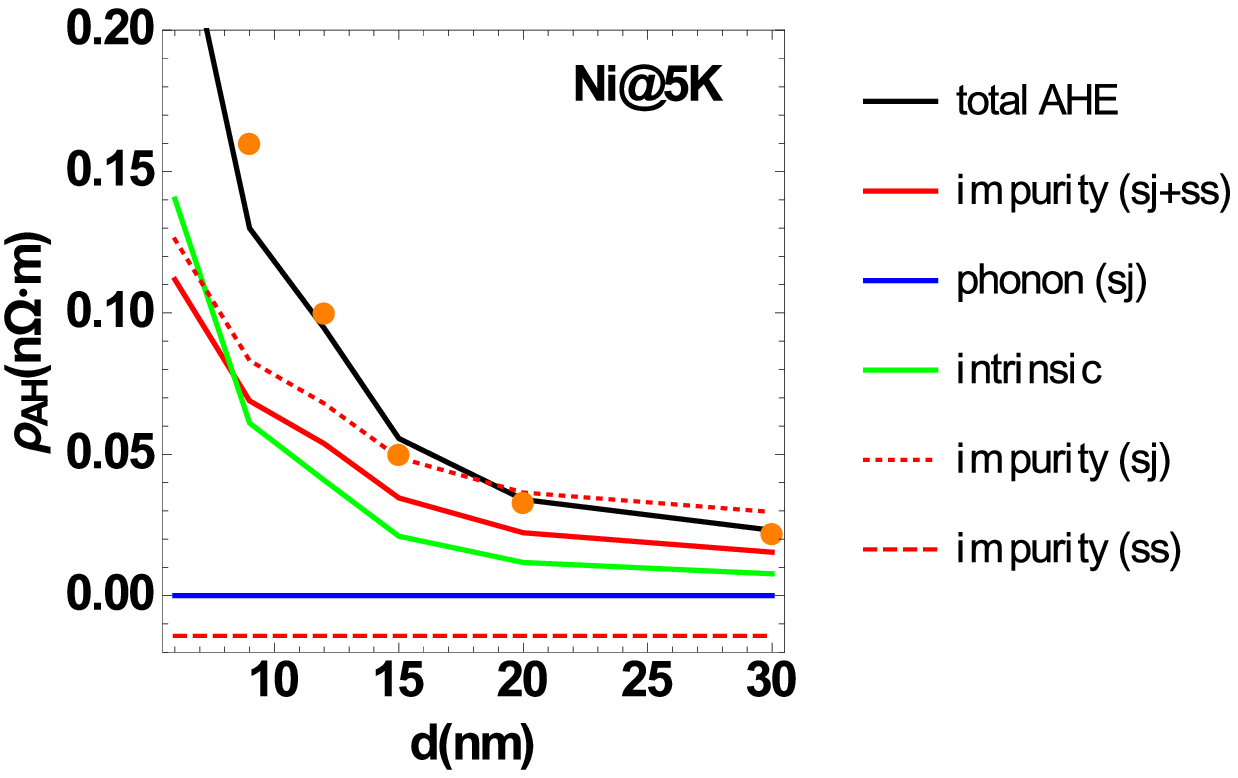}
\includegraphics[height=3.7cm,bb=20 10 230 230,clip]{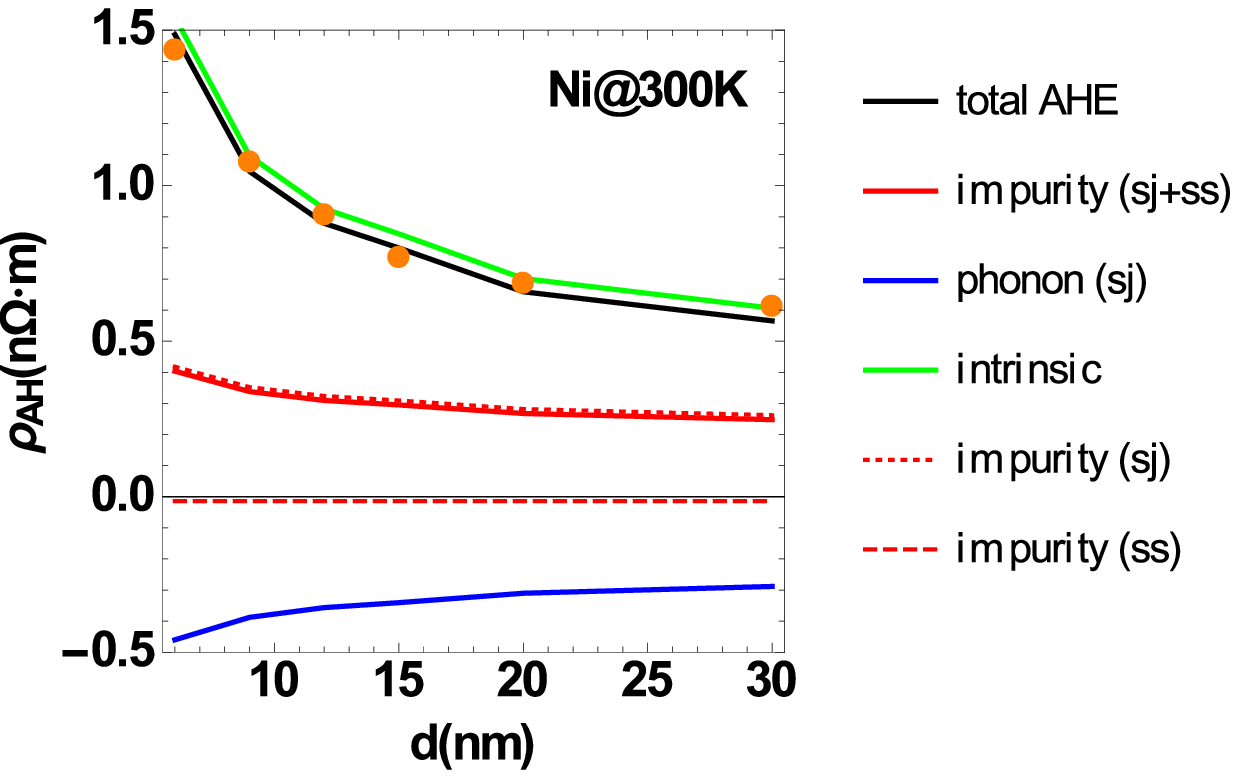}
\caption{(Color online) The same as in \Figure{fig:FeCo} for Ni. The points are the experimental data from Ref. \cite{ye_temperature_2012}.}\label{fig:FeNi}
\end{figure*}

We test our longitudinal resistivity \Eq{eqn:sigxx} using the experimental data obtained by the Jin's group for Fe, Co, and Ni \cite{hou_multivariable_2015, hou_anomalous_2012, ye_temperature_2012}, for which the film thickness and temperature are independently tuned. The only two free parameters in \Eq{eqn:sigxx} are the film surface roughness $\delta$ and (temperature-dependent) bulk relaxation time $\tau_0(T)$. We assume that $\delta$ is the same for all thicknesses, as the growth condition remains the same for all films. The relaxation time $\tau_0$ depends on temperature as it includes the phonon induced relaxation. Therefore only the data obtained at the same temperature has the same $\tau_0$. The fitting of thickness dependence of the longitudinal resistivity (\Eq{eqn:sigxx}) at different temperatures to the experimental data for Co is shown in \Figure{fig:rhoxxslices}. All fitting curves share the same roughness $\delta/a_0 = 5.7$. The inset shows the temperature dependence of the fitted values for $\tau_0$, which reflects the temperature dependence of the bulk (impurity and phonon induced) relaxation time. The full comparison of \Eq{eqn:sigxx} with experiment for Co is shown in Figures \ref{fig:FeCo}(a) (Figure \ref{fig:FeFe}(a) and \ref{fig:FeNi}(a) for Fe and Ni). The increase in the resistivity with decreasing film thickness is due to enhanced scattering from the surface roughness \cite{tesanovic_quantum_1986,zhou_spin_2015}.

\begin{figure*}[t]
\includegraphics[width=0.60\textwidth,bb=10 0 292 22 ,clip]{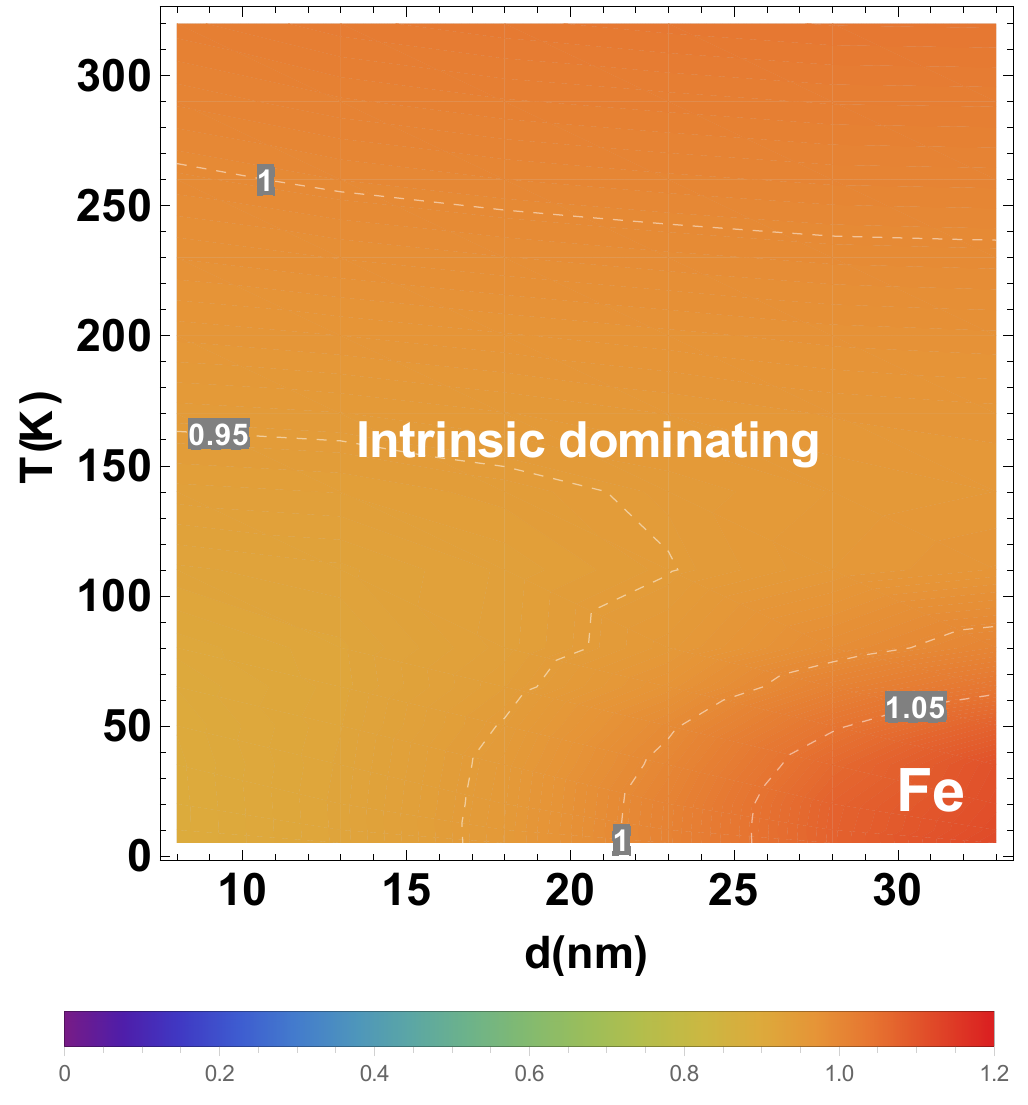}\\
\includegraphics[width=0.30\textwidth,bb=0 30 300 320,clip]{fig/int_Fe2.pdf}\hspace{0.5cm}
\includegraphics[width=0.30\textwidth,bb=0 30 300 320,clip]{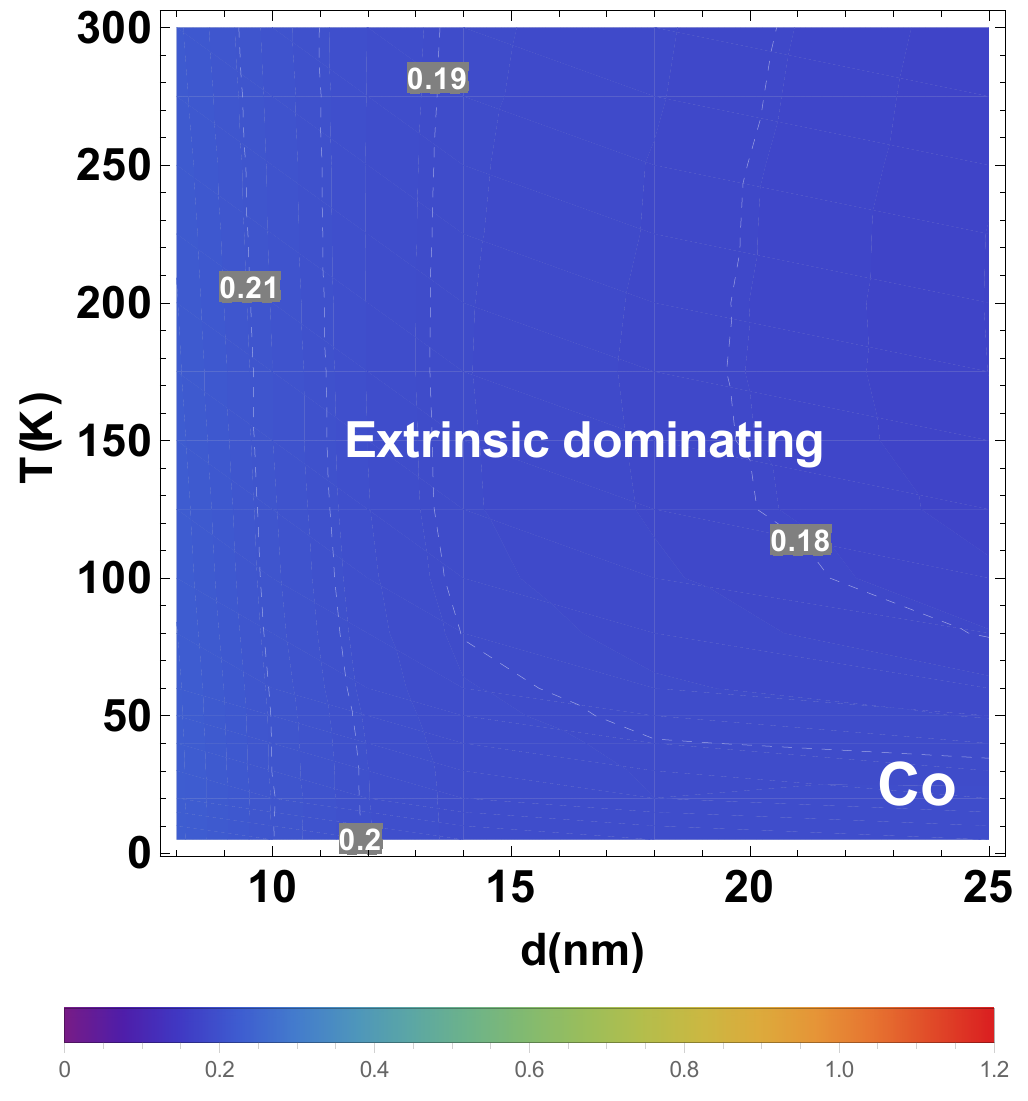}\hspace{0.5cm}
\includegraphics[width=0.30\textwidth,bb=0 30 300 320,clip]{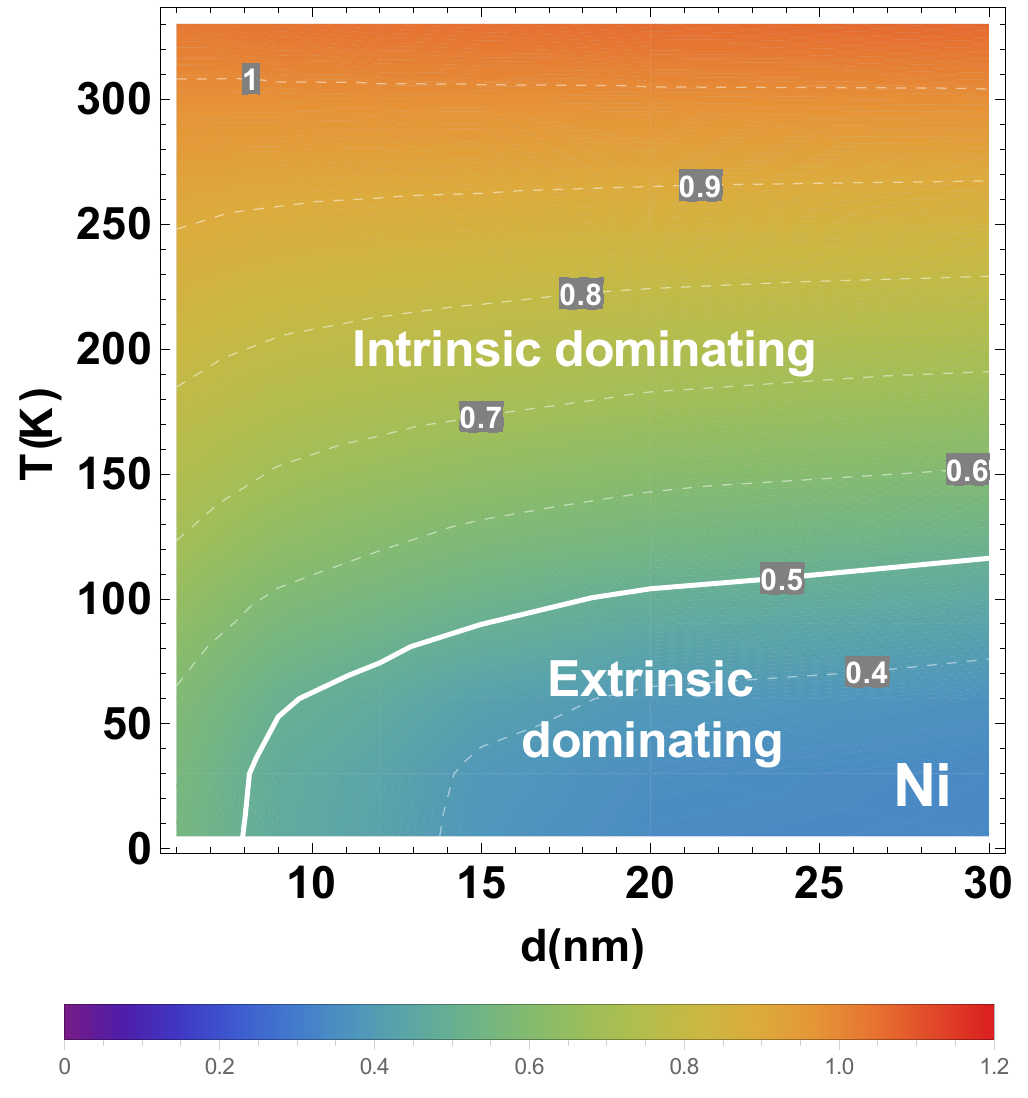}
\caption{(Color online) The ratio of intrinsic contribution to the total AHE: $r_\ssf{INT} = \rho_\ssf{AH}^\ssf{INT}/\rho_\ssf{AH}$ as function of film thickness and temperature for Fe, Co, Ni. For Fe, the AHE is dominated by the intrinsic mechanism with $r_\ssf{INT} > 90\%$. For Co, the AHE is dominated by extrinsic mechanisms with $r_\ssf{INT} \sim 20\%$. For Ni, the solid curve, corresponding to $r_\ssf{INT} = 50\%$, separates the intrinsic dominating region ($r_\ssf{INT} > 50\%$) from the extrinsic dominating region ($r_\ssf{INT} < 50\%$).}
\label{fig:int}
\end{figure*}

{\it Anomalous  Hall resistivity.} We now consider the transverse anomalous Hall conductivity in a ferromagnetic thin film with a magnetization perpendicular to the film plane. Several mechanisms contribute, including the intrinsic contribution from the band structure \cite{karplus_hall_1954} and the extrinsic contributions, i.e., side-jump and skew-scattering, originating from the spin-orbit coupling by the impurities and phonons. The intrinsic contribution has been well studied and is assumed to be independent of the geometry of the sample. The intrinsic anomalous Hall resistivity scales quadratically with the longitudinal resistivity: $\rho_\ssf{AH}^\ssf{INT} = \gamma \rho^2$ \cite{nagaosa_anomalous_2010}. Or, in terms of conductivities, it is $\sigma_\ssf{AH}^\ssf{INT} = \gamma$. Yet the extrinsic contributions depend on the detailed scattering processes.

Following the calculation of the spin Hall conductivity in normal metals \cite{zhou_spin_2015,wang_scaling_2014}, we can further obtain the anomalous conductivity in a ferromagnetic one. The conductivity due to side-jump is
\begin{equation}
\sigma_\ssf{AH}^\ssf{SJ}
= \alpha_\ssf{I} {\sigma \ov \sigma_\ssf{I}}
+ \alpha_\ssf{P} {\sigma \ov \sigma_\ssf{P}}
\qwith \alpha_\ssf{I,P}=\eta_\ssf{I,P} {n_ee^2\ov 2\hbar}{J_\ssf{sd}\ov \EF},
\label{eqn:sj}
\end{equation}
where $\eta_\ssf{I,P},\alpha_\ssf{I,P}$
 are the spin-orbit coupling constant and side-jump coefficient due to impurity/phonon scatterings, and $\sigma_\ssf{I,P} = n_ee^2\tau_\ssf{I,P}/m$ is the bulk Drude conductivity due to impurity/phonon relaxation alone. The skew scattering contribution is
\begin{equation}
\sigma_\ssf{AH}^\ssf{SS}
= \beta_\ssf{I} {\sigma^2\ov \sigma_\ssf{I}}
+ \beta_\ssf{P} {\sigma^2\ov \sigma_\ssf{P}}
\qwith \beta_\ssf{I,P} = -\eta_\ssf{I,P} {\pi n_e m V_\ssf{I,P}\ov
2\hbar^2}{J_\ssf{sd}\ov \EF}. \label{eqn:ss}
\end{equation}
where $\beta_\ssf{I,P}$ are skew-scattering coefficient due to impurity/phonon scatterings.  The overall anomalous Hall conductivity is the sum of all contributions
\begin{equation}
\sigma_\ssf{AH} 
= \alpha_\ssf{I}{\sigma\ov \sigma_\ssf{I}}+\beta_\ssf{I}{\sigma^2\ov \sigma_\ssf{I}}
+ \alpha_\ssf{P}{\sigma\ov \sigma_\ssf{P}}+\beta_\ssf{P}{\sigma^2\ov \sigma_\ssf{P}}
+\gamma,
\end{equation}
and when expressed in terms of resistivity,
\begin{align}
\rho_\ssf{AH} 
= {\sigma_\ssf{AH}\ov \sigma^2}
&=\alpha_\ssf{I}\rho_\ssf{I}\rho+\beta_\ssf{I}\rho_\ssf{I}
 +\alpha_\ssf{P}\rho_\ssf{P}\rho+\beta_\ssf{P}\rho_\ssf{P}
+\gamma\rho^2. 
\label{eqn:scaling}
\end{align}
In the above $\rho(T,d,\delta)$ is the longitudinal resistivity of the thin film and $\rho_\ssf{I,P} = \sigma_\ssf{I,P}^{-1}$ is the bulk limit of the Drude resistivity with only the impurity/phonon relaxation. Parameters $\alpha_\ssf{I,P}, \beta_\ssf{I,P}, \gamma$ are material constants that are independent of the film thickness/roughness and temperature. As the central result of this Letter, \Eq{eqn:scaling} provides the scaling law for the anomalous Hall effect in a ferromagnetic thin film with constant coefficients.

The new scaling law \Eq{eqn:scaling} is tested using the experimental data from Jin's group. We first extract the impurity induced skew-scattering parameter $\beta_\ssf{I}$ in \Eq{eqn:scaling} by fitting to the experimental data with varying thickness but at low temperatures ($\rho_\ssf{P}\simeq 0$), at which \Eq{eqn:scaling} reduces to a quadratic function of $\rho$, and the interception at $\rho = 0$ is $\beta_\ssf{I}\rho_\ssf{I}$. Due to the fluctuating sign of the effective potential $V_\ssf{P}$ from phonon scattering, the corresponding skew-scattering involving the odd power ensemble average $\avg{V_\ssf{P}^3}$ is negligible \cite{phonon_skew_2}, we thus set $\beta_\ssf{P} = 0$. With $\beta_\ssf{I,P}$ fixed, the full experimental data is fitted using \Eq{eqn:scaling} with $\alpha_\ssf{I, P}$ and $\gamma$ as fitting parameters. The fitted values for $\alpha_\ssf{I,P}, \beta_\ssf{I, P}$, and $\gamma$ are listed in \Table{tab:param}. The comparison of the fitted $\rho_\ssf{AH}$ in \Eq{eqn:scaling} (surface) with the experimental data (points) for Co is shown in Figures \ref{fig:FeCo}(b) (Figure \ref{fig:FeFe} (b) and \ref{fig:FeNi} (b) for Fe and Ni). Apparently, the agreement is excellent.

 The thickness dependence of the various AHE contributions for Co is shown in Figs. \ref{fig:FeCo}(c,d) at low and room temperatures, from which we see that extrinsic contributions dominates in Co, regardless the temperature. \Figure{fig:int}(b) plots the ratio between the intrinsic contribution and the total AHE $r_\ssf{INT} = \rho_\ssf{AH}^\ssf{INT}/\rho_\ssf{AH}$, which is about $20\%$ of the total AHE for Co in the full temperature and film thickness range. This means that the extrinsic mechanisms dominates in Co. On the other hand, the intrinsic effect dominates in Fe with $r_\ssf{INT} > 90\%$ in the full range for Fe. For Ni, the intrinsic (extrinsic) effect dominates at high (low) temperatures.   

{\it Discussion.} Previous scaling laws \cite{tian_proper_2009,hou_multivariable_2015} are not only complex but also contingent to fitting parameters that are simultaneously dependent on temperature and thickness. Such drawback are usually a strong indicator that an understanding of fundamental physics is largely missing. In contrast, our scaling law \Eq{eqn:scaling} provides an excellent agreement with the experimental data. More importantly, the fitting parameters $\alpha_{\ssf{I,P}},$ $\beta_{\ssf{I}}$, and $\gamma$ are real constants that are indeed independent of film thickness or temperature. Therefore, \Eq{eqn:scaling} is simple and carries a clear physical picture, \ie separating the surface scattering from the bulk ones (impurity and phonon scattering), and is sufficient to understand both the longitudinal and transverse transport behavior in ferromagnetic thin films. With such a highly satisfied fitting to the experimental data using the new law \Eq{eqn:scaling}, we are able to - for the first time - accurately determine the weight of each mechanism. 



In conclusion, by incorporating the scatterings from surface roughness in a ferromagnetic thin film, we found that the anomalous Hall resistivity scales with the longitudinal resistivity as $\rho_\ssf{AH} =\alpha_\ssf{I}\rho_\ssf{I}\rho+\beta_\ssf{I}\rho_\ssf{I} +\alpha_\ssf{P}\rho_\ssf{P}\rho +\gamma\rho^2$. In Fe, Co, and Ni thin films, this simple yet elegant relation agrees excellently with the experimental data. We conclude that the intrinsic (extrinsic) mechanism dominates in Fe (Co). But in Ni, the relative importance of the intrinsic and extrinsic mechanisms depends on temperature and film thickness.

{\it Acknowledgements.} The authors thank Xiaofeng Jin of Fudan Unviersity for valuable discussions. This work was supported by the National Natural Science Foundation of China under Grant No. 11474065, National Basic Research Program of China under Grant No. 2014CB921600.

\bibliographystyle{apsrev}




\end{document}